\pdfoutput=1
\documentclass[11pt]{article}

\usepackage[]{acl}

\usepackage{times}
\usepackage{latexsym}

\usepackage[T1]{fontenc}
\usepackage[utf8]{inputenc}

\usepackage{microtype}

\usepackage{inconsolata}

\usepackage{enumitem}
\usepackage{url}            % simple URL typesetting
\usepackage{graphicx} % DO NOT CHANGE THIS
\usepackage{xcolor}         % colors
\usepackage{subcaption}
\usepackage{booktabs}       % professional-quality tables
\usepackage{tabularx}
\usepackage{multirow}
\usepackage{amsmath}
\usepackage{amsfonts}       % blackboard math symbols
\usepackage{longtable}

\usepackage{pifont}% http://ctan.org/pkg/pifont
\newcommand{\cmark}{\ding{51}}%
\newcommand{\xmark}{\ding{55}}%

\newcommand{\ie}{\emph{i.e., }}

\newcommand{\Mat}[1]{\textbf{#1}}

\newcommand{\Space}[1]{\mathbb{#1}}

\newcommand{\Vtr}[1]{\boldsymbol{#1}}

\title{ProtT3: Protein-to-Text Generation for Text-based Protein Understanding}

\author{Zhiyuan Liu$^\dag$ \quad An Zhang$^\dag$ \quad Hao Fei$^\dag$ \quad Enzhi Zhang$^\S$ \\ \textbf{Xiang Wang$^\ddag$\thanks{~~Corresponding author. Xiang Wang is also affiliated with Institute of Dataspace, Hefei Comprehensive National Science Center.} \quad Kenji Kawaguchi$^\dag$ \quad Tat-Seng Chua$^\dag$}\\
$^\dag$National University of Singapore \\
$^\ddag$University of Science and Technology of China \quad 
$^\S$Hokkaido University\\
\texttt{\{acharkq,an.zhang3.14,xiangwang1223\}@gmail.com}\\ \texttt{enzhi.zhang.n6@elms.hokudai.ac.jp}, \texttt{\{feih, kenji,chuats\}@comp.nus.edu.sg}\\}

\begin{document}
\maketitle

% idea: we have two types of language models: 1) LMs are good at texts but not proteins; 2) PLMs are good at proteins but not texts. we need to connect these two models to utilize their strengths.

\begin{abstract}
    % lms are strong
    % Language Models (LMs) excel in understanding texts describing proteins, such as in biomedical question-answering tasks.
    % lms have an limitation
    % However, they lack the ability to understand raw protein data like sequences of amino acids, due to the lack of pretraining on such data. Conversely, Protein Language Models (PLMs) can generate effective protein representations but cannot process texts.
    Language Models (LMs) excel in understanding textual descriptions of proteins, as evident in biomedical question-answering tasks. However, their capability falters with raw protein data, such as amino acid sequences, due to a deficit in pretraining on such data. Conversely, Protein Language Models (PLMs) can understand and convert protein data into high-quality representations,  but struggle to process texts.
    % we propose a model to address the limitation
    To address their limitations, we introduce \textbf{ProtT3}, a framework for \underline{Prot}ein-\underline{t}o-\underline{T}ext Generation for \underline{T}ext-based Protein Understanding. 
    % what does your model do?
    ProtT3 empowers an LM to understand protein sequences of amino acids by incorporating a PLM as its protein understanding module, enabling effective protein-to-text generation. 
    This collaboration between PLM and LM is facilitated by a cross-modal projector (\ie Q-Former) that bridges the modality gap between the PLM's representation space and the LM's input space. 
    % how does ProtT3 do this?
    Unlike previous studies focusing on protein property prediction and protein-text retrieval, we delve into the largely unexplored field of protein-to-text generation. To facilitate comprehensive benchmarks and promote future research, we establish quantitative evaluations for protein-text modeling tasks, including protein captioning, protein question-answering, and protein-text retrieval. Our experiments show that ProtT3 substantially surpasses current baselines, with ablation studies further highlighting the efficacy of its core components. Our code is available at \url{https://github.com/acharkq/ProtT3}.
\end{abstract}
\section{Introduction}
% Level1: It is urgent to use language model for biology research 
Language Models (LMs) have achieved impressive successes across diverse domains~\cite{GPT3, Llama2, liu2023molca}. Remarkably, the extensive biological literature in their training data has enabled LMs to excel in text-based protein understanding tasks, such as biological and medical question-answering (QA)~\cite{Galactica,GPT4}. These results show the potential of using LMs as the natural language interface for biomedical tasks. It further accentuates the importance of harnessing LMs to drive advancements in areas like drug discovery and protein property prediction~\cite{kim2021comprehensive,CDConv}.

% emphasizing the necessity to exploit LMs to accelerate biology and medical research, such as  and 

% \begin{figure}[t]
% \centering
% \small
% \begin{subfigure}[b]{0.42\textwidth}
% \includegraphics[width=\textwidth]{figures/protein_captioningv2}
% % \includegraphics[width=\textwidth]{figures/protein_captioningv2}
% \caption{Protein captioning.}
% \label{fig:protein_captioning}
% \end{subfigure}
% \begin{subfigure}[b]{0.42\textwidth}
% \centering
% \vspace{2mm}
% \includegraphics[width=\textwidth]{figures/protein_qav2}
% % \includegraphics[width=\textwidth]{figures/protein_qav2}
% \caption{Protein question-answering.}
% \label{fig:protein_qa}
% \end{subfigure}
% \vspace{-2mm}
% \caption{Examples of protein-to-text generation tasks. Proteins are represented by sequences of amino acids.}
% \label{fig:examples}
% \end{figure}

\begin{figure}[t]
\centering
\small
\includegraphics[width=0.44\textwidth]{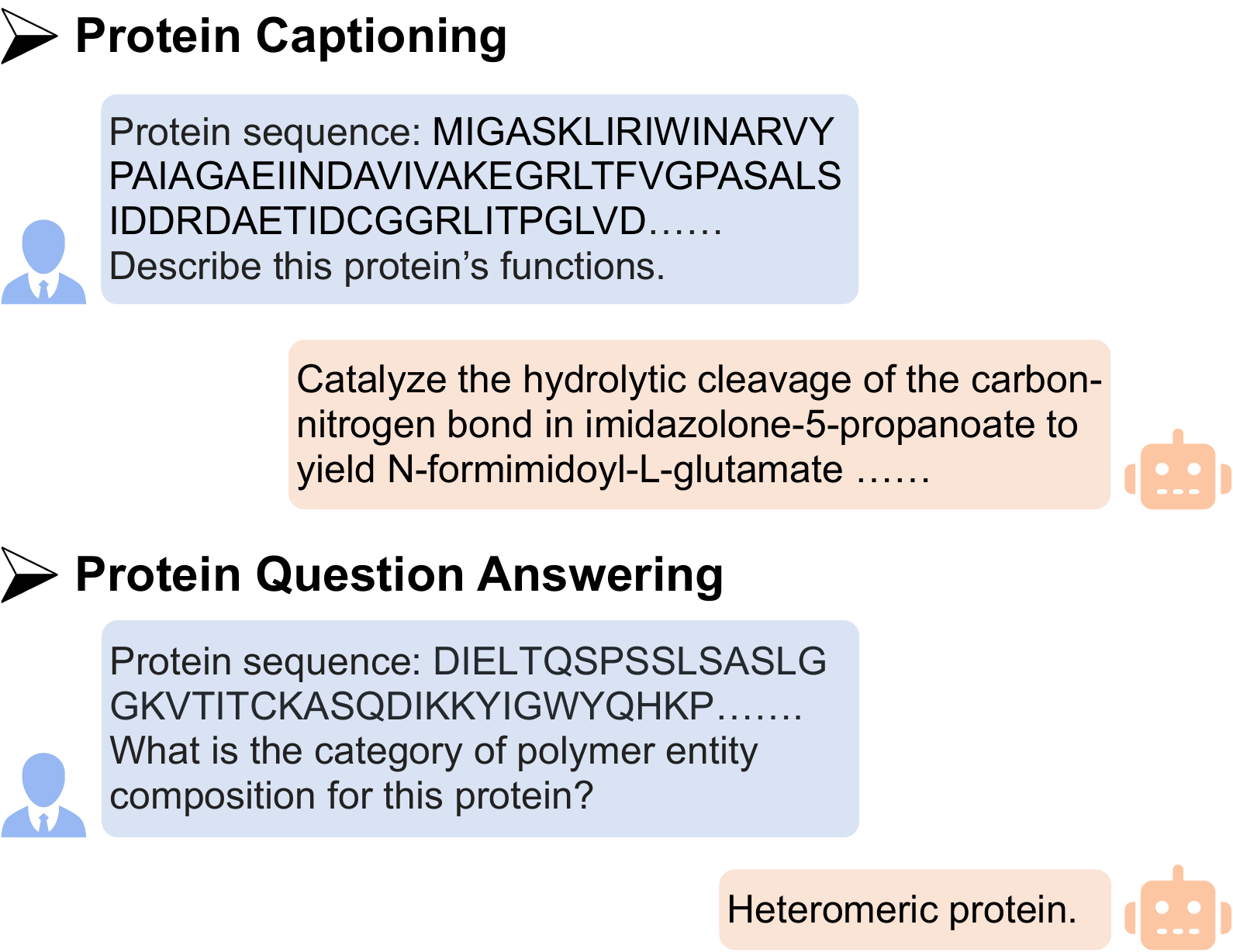}
\caption{Examples of protein-to-text generation tasks. Proteins are represented by sequences of amino acids.}
\label{fig:protein_to_text}
\vspace{-0.2cm}
\end{figure}

\begin{table}[t]
\small
\centering
\setlength{\tabcolsep}{0.8pt}
\begin{tabular}{lccc} \toprule
             & \multicolumn{2}{c}{P2T Generation} & \multirow{2}{*}{\shortstack{ \\ Prot-Text \\ Retrieval}}   \\\cmidrule{2-3}
Methods      & Caption              & QA                      \\ \midrule
ProteinCLAP~\cite{ProteinDT}  & $\text{\xmark}$         & $\text{\xmark}$        & {\large \textbf{?}}        \\
ProtST~\cite{ProtST}      & $\text{\xmark}$         & $\text{\xmark}$        & {\large \textbf{?}}        \\
Galactica~\cite{Galactica}    & $\text{\cmark}$   & {\large \textbf{?}}  & $\text{\xmark}$        \\
ProteinChat~\cite{ProteinChat}  & {\large \textbf{?}}   & {\large \textbf{?}}  & $\text{\xmark}$        \\
Ours, ProtT3 & $\text{\cmark}$         & $\text{\cmark}$        & $\text{\cmark}$       \\\bottomrule
\end{tabular}
\caption{Comparing the abilities of protein-text modeling methods. {\large \textbf{?}} denotes missing quantitative evaluation.}
\vspace{-6mm}
\label{tab:previous_works}
\end{table}

\begin{figure*}
\centering
\small
% \begin{minipage}{.5\textwidth}
\begin{subfigure}[t]{0.59\textwidth}
\includegraphics[width=\textwidth]{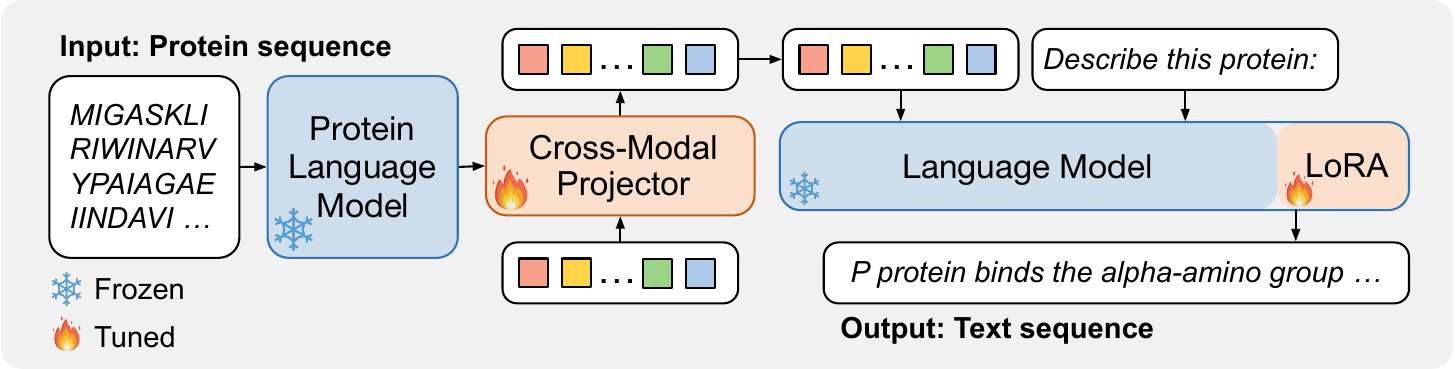}
\caption{ProtT3's architecture.}
\label{fig:framework_arch}
\end{subfigure}
\hspace{2pt}
% \begin{minipage}{.35\textwidth}
\begin{subfigure}[t]{0.38\textwidth}
\includegraphics[width=\textwidth]{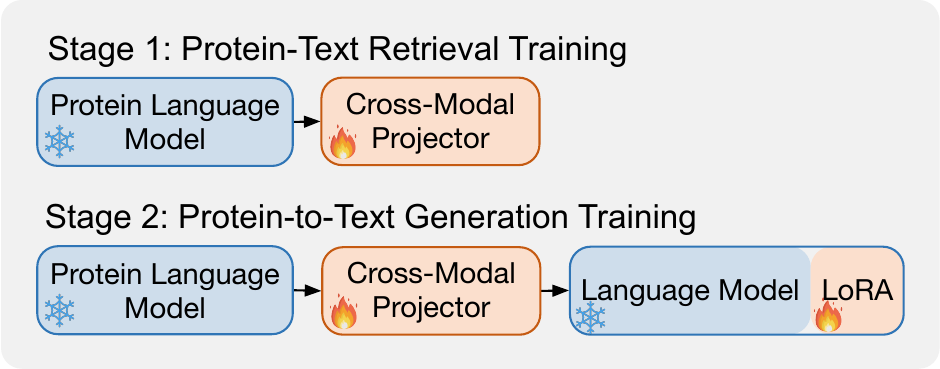}
\caption{ProtT3's two training stages.}
\label{fig:framework_stage}
\end{subfigure}
\caption{Overview of the ProtT3 framework.}
\label{fig:prott3}
\vspace{-4mm}
\end{figure*}

Here we focus on LM's capability in text-based protein understanding, which enables textual interpretations of proteins. It can be assessed through diverse downstream tasks, particularly protein-to-text generation~\cite{ProteinChat} and retrieval~\cite{ProtST}. Specifically, the retrieval task is finding an existing text description best matching a particular protein, while the generation task is delineated into two problems, as illustrated in Figure~\ref{fig:protein_to_text}:
1) protein captioning, where the LM generates a description of a specific protein's functions, and 2) protein QA, where the LM answers questions about a protein.
% Toward this end, we focus on protein-to-text generation and protein-text retrieval, aiming to faicilitate text-based protein understanding. As illustrated in Figure~\ref{fig:examples}, a protein-to-text generation system \yuan{should be able} to describe the protein's functions (\ie protein captioning) or answer questions about protein properties (\ie protein QA). 
Table~\ref{tab:previous_works} overviews prior efforts~\cite{ProteinDT, ProtST,Galactica, ProteinChat} in this field. Delving into these previous studies, we identify two research gaps:
\begin{itemize}[leftmargin=*]
\item \textbf{Lack of Exploration for Protein-to-Text Generation.} 
Protein-to-text generation is a conditional generation task~\cite{keskar2019ctrl}, requiring the LM to perceive proteins as the generation condition. 
Previous studies~\cite{ProtST, ProteinDT} based on cross-modal contrastive learning~\cite{CLIP} hardly interpret proteins as the direct inputs to the LM.
% basically fails, as the proteins are not employed as direct inputs to the LM.
% are insufficient for this purpose for not using proteins as the LM's input. 
% \yuan{The performances of two other works, Galactica and ProteinChat, can be considered suboptimal for different reasons: 
While notably two studies, Galactica and ProteinChat, have paid certain explorations, they unfortunately appear to be quite constrained by key limitations.  
Specifically, Galactica~\cite{Galactica} incorporates only a limited set of protein sequences in its pretraining data, thereby potentially restricting its capacity for comprehensive protein understanding;
% includes only a small set of protein sequences in its pretraining data, thus easily limited in protein understanding; 
ProteinChat~\cite{ProteinChat} seeks  to project protein representations to text space by training a linear projector, which might prove inadequate in capturing the intricate relations between proteins and texts.
% \item \textbf{Missing Quantitative Evaluation.} Tracking the progress in protein-text modeling is often challenging in the absence of benchmarks, as the quantitative evaluations of these models are mostly missing, as shown in Table~\ref{tab:previous_works}.
% This is because previous studies either focus on other tasks or include only case studies. 
% \yuan{This limitation will hinder making progress  text-based protein understanding.}
\item \textbf{Missing Quantitative Evaluation.} The progress in protein-text modeling is difficult to track without proper benchmarks. As Table~\ref{tab:previous_works} illustrates, the quantitative evaluations for these models are mostly missing, posing a challenge to further advancement in this field.
% the adoption and improvement of previous studies in the research community.
\end{itemize}

To bridge these research gaps, we propose \textbf{ProtT3}: \underline{Prot}ein-\underline{t}o-\underline{T}ext Generation for \underline{T}ext-based Protein Understanding. As Figure~\ref{fig:framework_arch} illustrates, ProtT3 empowers an LM to understand protein sequences of amino acids by incorporating a Protein Language Model (PLM) as its protein understanding module, thereby effectively conditioning the protein-to-text generation process. PLMs~\cite{ProGen} are specialized LMs pretrained solely on protein sequences. They can generate powerful protein representations that are instrumental in presenting the proteins' 3D structures and indicating their potential properties~\cite{xTrimoPGLM}. To enable the LM to understand the PLM's protein representations, ProtT3 integrates an expressive cross-modal projector -- Q-Former~\cite{BLIP2} -- to map protein representations into the text space of the LM. This design enables the LM to consume proteins as inputs. However, working with a large LM with billions of parameters raises a new challenge of maintaining the efficiency of downstream adaptation. Therefore, we incorporate a LoRA~\cite{LoRA} adapter into the LM for efficient fine-tuning purposes.

% ProtT3 jointly leverages an LM for text processing and a Protein Language Model (PLM) to encode protein sequences of amino acids. PLMs~\cite{ProGen} are specialized LMs pretrained solely on protein sequences. They can generate powerful protein representations that are instrumental in presenting the proteins' 3D structures and indicating their potential properties~\cite{ESM2,xTrimoPGLM}. To enable the LM to understand the PLM's protein representations, ProtT3 integrates an expressive cross-modal projector -- Q-Former~\cite{BLIP2} -- to map protein representations into the text space of the LM. This design enables the LM to consume proteins as inputs, thereby effectively conditioning the protein-to-text generation process. However, working with a large LM with billions of parameters raises a new challenge of maintaining the efficiency of downstream adaptation. Therefore, we incorporate a LoRA~\cite{LoRA} adapter into the LM for efficient fine-tuning purposes.

To facilitate effective protein-text modeling, ProtT3 employs a two-stage training process to enhance protein-text modeling, as outlined in Figure~\ref{fig:framework_stage}. The first stage involves protein-text retrieval training with three cross-modal tasks~\cite{BLIP}: protein-text contrasting, protein-text matching, and protein captioning. This stage not only empowers the cross-modal projector with the capability of protein-text retrieval, but also serves as a ``warmup'' before the second stage by encouraging the extraction of text-relevant protein features. In the second stage, we connect the cross-modal projector to the LM and conduct protein-to-text generation training. 

Our key contributions are summarized as:
\begin{itemize}[leftmargin=*]
    \item We introduce ProtT3, a new framework aiming to bridge the modality gap between texts and proteins. Through a cross-modal projector, ProtT3 jointly uses a PLM for protein understanding and an LM for text processing, enabling effective protein-to-text generation.
    % \yuan{Further, we establish quantitative evaluations for protein-to-text generation tasks, paving the way for future research.}
    \item To set benchmarks and promote future research, we establish quantitative evaluations for protein-text modeling tasks, including protein captioning, protein QA, and protein-text retrieval. The datasets, evaluation scripts, and our pretrained checkpoints will be made available online.
    \item ProtT3 achieves state-of-the-art performances across various tasks. For protein captioning, ProtT3 surpasses the baseline by over 10 BLEU-2 scores in the Swiss-Prot~\cite{SwissProt} and the ProteinKG25~\cite{OntoProtein} datasets. For protein-text retrieval, ProtT3 outperforms baselines by over $14\%$ in retrieval accuracy on the Swiss-Prot and ProteinKG25 datasets. Lastly, ProtT3 achieves 2.5\% improvement of exact match performance for protein QA on the PDB-QA~\cite{ProteinChat} dataset.
    % \item To support future research, we will make our codes, models, and datasets available online.
    % \item ablation studies show the efficacy of ProtT3's core components.
\end{itemize}

\section{Related Works}
Here we briefly review relevant fields of PLMs, protein-text modeling, and multi-modal LMs.

\textbf{Protein Language Models (PLMs).} PLMs are transformer-based LMs pretrained on large corpora of protein sequences for protein understanding and protein generation~\cite{ESM2, ProGen, ProGen2, xTrimoPGLM, ProtTrans, DBLP:journals/pnas/RivesMSGLLGOZMF21, DBLP:conf/nips/MeierRVLSR21}. Similar to LMs for texts, PLMs are pretrained by the objective of masked language modeling~\cite{ESM2} or auto-regressive modeling~\cite{ProGen}. PLMs have demonstrated promising performances on downstream tasks of 3D structure prediction and protein property prediction~\cite{ESM2, xTrimoPGLM}. However, they cannot process texts due to the absence of text in their pretraining data.

\begin{figure*}
\centering
\small
\includegraphics[width=0.95\textwidth]{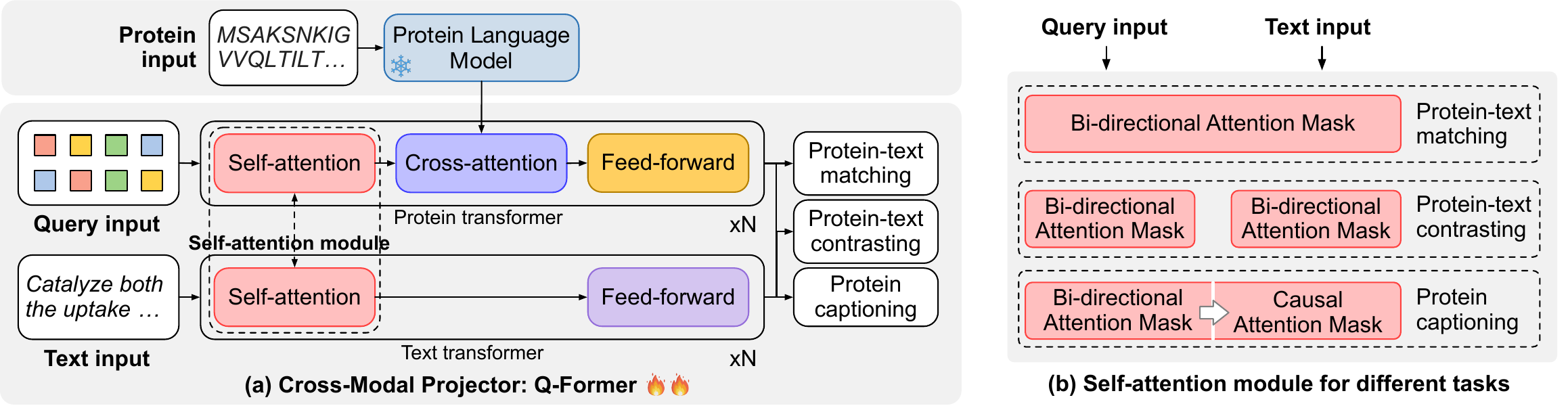}
\caption{The training stage 1 of ProtT3. \textbf{(a):} Cross-Model Projector:  Q-Former's architecture and the three training tasks. \textbf{(b):} The self-attention module uses different masking strategies for different tasks.}
\label{fig:qformer}
\vspace{-5mm}
\end{figure*}

\textbf{Protein-Text Modeling.} The pioneering study, Galactica~\cite{Galactica}, is an LM pretrained on a text corpus that includes a small amount of protein sequences. Its protein understanding ability can be suboptimal compared to PLMs due to insufficient protein pretraining data. To mitigate this gap, subsequent works~\cite{ProtST, ProteinDT} jointly leverage a PLM and an LM by cross-modal contrastive learning~\cite{CLIP}. However, cross-modal contrastive learning is insufficient for protein-to-text generation. This is because protein-to-text generation is a conditional generation task. It demands the LM to understand proteins as the generation condition, which contrastive learning cannot achieve. Further, ProteinChat~\cite{ProteinChat} attempts to enable an LM to understand proteins by training a linear projector between a PLM and an LM. Considering both the PLM and LM are kept frozen during training, this approach struggles to capture the nonlinear relations between proteins and texts. In a different approach, OntoProtein~\cite{OntoProtein} fine-tunes a PLM using a biology knowledge graph, focusing more on protein property prediction and less on text-related tasks.

\textbf{Multi-modal LMs.} Enabling LMs to understand another modality, such as image~\cite{BLIP, Frozen}, video~\cite{Video-LLaMA}, audio~\cite{lyu2023macawllm}, user-items~\cite{liao2024llara}, and molecule~\cite{liu2023molca, ReLM, 3D-MoLM, fang2024moltc, MolT5}, has been an activate research direction. Notably, the research of multi-modal LMs is pioneered by vision-language modeling (VLM). VLM has been successfully applied for few-shot image classification, image captioning, and image QA~\cite{BLIP,Flamingo}. Specifically, to enable LMs to understand images, the leading VLM methods employ either 1) nonlinear and expressive cross-modal projectors~\cite{BLIP2,Flamingo}, or 2) fine-tuned visual encoders and LMs on multi-modal datasets~\cite{PALM-E}, while such studies are missing for protein-text modeling. Our work takes the initiative to explore this direction.

% They have inspired similar works in the audio~\cite{lyu2023macawllm}, video~\cite{Video-LLaMA}, and molecules~\cite{MoMu, MolT5}. 

% \yuan{Parallel to protein-text modeling, multi-modal molecule-text modeling aims to bridge the modality gap between small molecules and texts. Following a similar trend, previous studies in this area have explored cross-modal contrastive learning~\cite{MoMu,MoleculeSTM,Text2Mol} and joint 1D sequence modeling of molecules and texts~\cite{MolT5, KVPLM}. }

% \begin{figure*}
% \centering
% \small
% \includegraphics[width=0.9\textwidth]{figures/framework1}
% \caption{Overview of the ProtT3 framework.}
% \label{fig:framework1}
% \end{figure*}

\vspace{-1mm}
\section{Model Architecture}
\vspace{-1mm}
Here we introduce ProtT3's three key components: a PLM for protein understanding, an LM for text processing, and a cross-modal projector to bridge the modality gap between the first two components.

\textbf{Protein Language Model (PLM).}
We utilize ESM-2~\cite{ESM2} to encode protein sequences of amino acids. ESM-2 is an encoder-only transformer LM~\cite{Transformer} pretrained on 60M protein sequences by masked language modeling~\cite{BERT}. PLMs have shown promising performances for protein folding~\cite{ESM2}, multiple sequence alignment~\cite{ESM-1b}, and protein property prediction~\cite{ProtST}, demonstrating their effectiveness for capturing protein characteristics. For efficiency, we freeze ESM-2's weights in our training process.

\textbf{Language Model (LM).} We choose Galactica~\cite{Galactica} as the base LM. Galactica is a decoder-only transformer LM pretrained on a large collection of scientific papers, spanning disciplines like biology and medicine. Notably, Galactica has demonstrated a high-level understanding of protein concepts through its promising performances in biomedical QA benchmarks~\cite{PubMedQA,MMLU}. Furthermore, Galactica includes a small set of protein sequences in its pretraining data, and shows a capability for understanding protein sequences through the task of protein keyword prediction~\cite{Galactica}. Therefore, we also leverage Galactica as a baseline for protein-to-text generation to ablate the effectiveness of incorporating an additional PLM for protein understanding.

\textbf{Cross-Modal Projector.}  We employ a cross-modal projector based on Q-Former~\cite{BLIP2} to bridge the modality gap between the PLM and the LM. Q-Former has demonstrated promising performances in vision-language tasks. As Figure~\ref{fig:qformer}a illustrates, Q-Former consists of two transformers: one for protein encoding and another for text processing.  
Specifically, the protein transformer maintains $N_q$ learnable query tokens $\{\Vtr{q}_i \in \Space{R}^d \}_{i=1}^{N_q}$ as inputs. These query tokens can interact with the PLM through the cross-attention modules, in order to extract protein features. We denote the protein transformer's output as $\Mat{Z} \in \Space{R}^{N_q \times d}$, containing protein features. For text input, the text transformer adds a \texttt{[CLS]} token at the beginning, and uses the \texttt{[CLS]} token's output as the text representation. Both transformers share the self-attention to enable interactions between proteins and texts. Details are in Section~\ref{sec:retrieval}. 

Q-Former's weights are initialized from PubMedBERT-Abstract~\cite{PubMedBERT}, a BERT LM pretrained on paper abstracts from the PubMed database\footnote{\url{https://pubmed.ncbi.nlm.nih.gov/}}. It has shown promising performances in understanding biomedical concepts under the BLURB benchmark~\cite{PubMedBERT}. The cross-attention module is added into the Q-Former every two layers and is randomly initialized.

\section{Training Method}
In this section, we introduce ProtT3's two training stages: protein-text retrieval and protein-to-text generation. The training process leverages a dataset of protein-text pairs $\mathcal{D}=\{(\Vtr{p}_1, \Vtr{t}_1), (\Vtr{p}_2, \Vtr{t}_2), ...\}$, where $\Vtr{p}_i$ is a protein sequence and $\Vtr{t}_i$ is the corresponding text sequence.

\subsection{Stage 1: Protein-Text Retrieval Training}
\label{sec:retrieval}
% In this and the next section, we utilize a dataset of protein-text pairs for training the ProtT3 framework.
Inspired by BLIP~\cite{BLIP, BLIP2}, we jointly employ three objectives for protein-text retrieval training: protein-text contrasting, protein-text matching, and protein captioning. These objectives are tailored for Q-Former's architecture, training it to extract protein features that are relevant to the text descriptions. This stage empowers the cross-modal projector with retrieval ability, and also serves as a ``warmup'' before the next stage.

\textbf{Protein-Text Contrasting (PTC).} We employ cross-modal contrastive learning~\cite{CLIP} to align the protein representation and text representation from the Q-Former. As illustrated in Figure~\ref{fig:qformer}b, Q-Former's self-attention module separately processes the query tokens and text tokens without any interaction. This enforces the query tokens to extract protein features from the PLM, in order to generate protein representations that align with the corresponding text representations in contrastive learning.

Formally, let $\{(\Vtr{p}_1, \Vtr{t}_1), ... (\Vtr{p}_B, \Vtr{t}_B)\}$ be a batch of protein-text pairs. We denote the protein representations as $\{\Mat{Z}_i\in \Space{R}^{N_q\times d}\}_{i=1}^{B}$, where $\Mat{Z}_{ij}\in \Space{R}^d$ is the reprentation of the protein $\Vtr{p}_i$'s $j$-th query token; and denote text $\Vtr{t}_i$'s representation as $\Vtr{m}_i \in \Space{R}^d$, which is the \texttt{[CLS]} token's output.
Protein-text similarity is measured by the maximum similarity between $\Vtr{m}$ and each row of $\Mat{Z}$. The contrastive learning loss $\mathcal{L}_{\text{PTC}}$ can then be written as:
{\small
\begin{gather}
    \nonumber \mathcal{L}_{\text{p2t}} = \frac{1}{B} \sum_{i=1}^B \log \frac{\exp(\max_k\cos(\Mat{Z}_{ik}, \Vtr{m}_i)/\tau)}{\sum_{j=1}^B \exp(\max_k \cos(\Mat{Z}_{ik}, \Vtr{m}_j)/\tau)}, \\
    \nonumber \mathcal{L}_{\text{t2p}} = \frac{1}{B} \sum_{i=1}^B \log \frac{\exp(\max_k\cos(\Mat{Z}_{ik}, \Vtr{m}_i)/\tau)}{\sum_{j=1}^B \exp(\max_k \cos(\Mat{Z}_{jk}, \Vtr{m}_i)/\tau)}, \\
    \mathcal{L}_{\text{PTC}} =  -\mathcal{L}_{\text{p2t}} - \mathcal{L}_{\text{t2p}},
\end{gather}}where $\cos(\cdot,\cdot)$ is the cosine similarity; Temperature $\tau$ is empirically set to $0.1$.

\textbf{Protein-Text Matching (PTM).} PTM is a binary classification objective aiming to discriminate whether a protein-text pair matches or not. Unlike PTC which computes protein-text similarity by applying cosine similarity on their output representations, PTM can obtain more fine-grained protein-text similarity. As illustrated in Figure~\ref{fig:qformer}b, PTM feeds the query tokens and text tokens into the same self-attention module, allowing them to interact at the Q-Former's early layers. In this way, the query tokens can capture information on both proteins and texts. The mean pooling of query tokens' representations is then fed into a linear classifier for PTM prediction. During training, for each positive protein-text pair $(\Vtr{p}_i, \Vtr{t}_i)$, we randomly sample two negative pairs: $(\Vtr{p}_i, \Vtr{t}_{i'})$ and $(\Vtr{p}_{i''}, \Vtr{t}_i)$. Let $f(\Vtr{p}, \Vtr{t})\in [0,1]$ be the Q-Former's predicted probability that $(\Vtr{p}, \Vtr{t})$ is a matched pair, the PTM loss function $\mathcal{L}_{\text{PTM}}$ can be written as:
\begin{gather}
    \nonumber \mathcal{L}_{\text{PTC}} = \frac{1}{B}\sum_{i=1}^B [-\log f(\Vtr{p}_i, \Vtr{t}_i) + \log f(\Vtr{p}_{i'}, \Vtr{t}_i) \\ + \log f(\Vtr{p}_{i}, \Vtr{t}_{i''})].
\end{gather}

In experiments, we find that PTM surpasses PTC for protein-text retrieval. However, PTM incurs a higher computational cost for encoding every protein-text pair. To balance performance and speed, we use PTC to obtain the top $k$ ranked candidates, and then use PTM for re-ranking.

\textbf{Protein Captioning (PCap).} Protein captioning trains the Q-Former to generate text descriptions of given proteins. As shown in Figure~\ref{fig:qformer}b, we apply a special masking strategy for this objective: 1) Bi-directional attention mask is applied to query tokens, enabling them to interact with each other but not text tokens; 2) Causal attention mask is used for text tokens, allowing them to attend query tokens and the preceding text tokens, but not the following text tokens. This design ensures the text tokens extract protein features exclusively from the query tokens, because they cannot directly interact with the proteins. Meanwhile, the query tokens are enforced to extract protein features through cross-attentions, satisfying the informational needs of protein captioning. Let $P(\Vtr{t}|\Vtr{p})$ be the probability that Q-Former generate text $\Vtr{t}$ given protein $\Vtr{p}$. The protein captioning loss $\mathcal{L}_{\text{PCap}}$ can be written as:
\begin{equation}
    \mathcal{L}_{\text{PCap}} = -\frac{1}{B}\sum_{i=1}^B \log P(\Vtr{t}_i|\Vtr{p}_i).
\end{equation}

\begin{table}[t]
\centering
\small
\begin{subtable}[t]{0.5\textwidth}
\centering
\small
\begin{tabular}{lccc}\toprule
    & Size   & Avg Protein Len & Avg Text Len \\\midrule
Train & 430595 & 336             & 48           \\
Valid & 10000  & 358             & 59           \\
Test  & 10000  & 357             & 60          \\\bottomrule
\end{tabular}
\caption{Swiss-Prot dataset of proteins and their text descriptions.}
\end{subtable}
\begin{subtable}[t]{0.5\textwidth}
\centering
\small
\begin{tabular}{lccc}\toprule
    & Size   & Avg Protein Len & Avg Text Len \\\midrule
Train & 422315 & 338             & 101          \\
Valid & 10000  & 360             & 104          \\
Test  & 10000  & 360             & 107         \\\bottomrule
\end{tabular}
\caption{ProteinKG25 dataset of proteins and their text descriptions.}
\end{subtable}
\begin{subtable}[t]{0.5\textwidth}
\centering
\small
\setlength{\tabcolsep}{3pt}
\begin{tabular}{lcccc} \toprule
    & \#Protein & \#QA Pair & Avg Protein Len & Avg Text Len \\\midrule
Train & 114690    & 3359693   & 291             & 10           \\
Valid & 10000     & 291795    & 257             & 10           \\
Test  & 10000     & 291779    & 259             & 10          \\\bottomrule
\end{tabular}
\caption{PDB-QA dataset. Each protein is acompanied by multiple QA pairs, therefore proteins and QA pairs have different numbers.}
\end{subtable}
\vspace{-2mm}
\caption{Statistics of our used protein-text datasets. Text lengths are counted by splitting at spaces.}
\label{tab:statistics}
\vspace{-4mm}
\end{table}

\begin{table*}[t]
\small
\centering
\begin{tabular}{p{10cm}|c} \toprule
\bf QA   Pair                                                                                  &\bf Type                      \\\midrule
Q: Does this protein   contain polymer entities? A: Yes                                   & String structure/property \\ \midrule
Q: What are the software programs reported in connection with the production of this protein? A: CNS, REFMAC, SCALEPACK & String supplementary information   \\ \midrule
Q: How many polymer   monomers does this protein have? A: 891                              & Number structure/property \\ \midrule
Q: When is this   protein first published? A: 2005                                         & Number supplementary information   \\\bottomrule
\end{tabular}
\caption{Sampled QA pairs from the PDB-QA dataset. The sampled protein's PDB id is 1xf2.}
\label{tab:sample}
\end{table*}

\begin{table*}[t]
\small
\centering
\setlength{\tabcolsep}{5pt}
\begin{subtable}[t]{\textwidth}
\centering
\small
\begin{tabular}{lc|ccccccc} \toprule
Model   & Base LM                      & Exact Match    & BLEU-2         & BLEU-4         & ROUGE-1        & ROUGE-2        & ROUGE-L        & METEOR         \\\midrule
LoRA fine-tune & $\text{Galac}_{\text{1.3B}}$ & 13.49          & 42.48          & 37.79          & 44.42          & 34.73          & 42.46          & 48.27          \\
ProtT3 w/ MLP Proj.      & $\text{Galac}_{\text{1.3B}}$ & 20.60                & 48.96           & 44.59           & 57.28            & 50.17          & 56.89          & 57.30          \\
ProtT3 w/o stage 1 & $\text{Galac}_{\text{1.3B}}$ & 22.88          & 50.21          & 46.76          & 58.64          & 51.63          & 57.17                & 58.62                \\
ProtT3  & $\text{Galac}_{\text{1.3B}}$ & \textbf{25.74} & \textbf{55.03} & \textbf{51.47} & \textbf{63.67} & \textbf{56.59} & \textbf{62.16} & \textbf{63.63}\\\bottomrule
\end{tabular}
\caption{Performances (\%) on the Swiss-Prot~\cite{SwissProt} dataset.}
\label{tab:swissprot}
\end{subtable}
\begin{subtable}[t]{\textwidth}
\centering
\small
\begin{tabular}{lc|ccccccc}\toprule
Model          & Base LM                      & Exact Match   & BLEU-2         & BLEU-4         & ROUGE-1        & ROUGE-2        & ROUGE-L        & METEOR         \\ \midrule
LoRA fine-tune & $\text{Galac}_{\text{1.3B}}$ & 2.67          & 64.97          & 56.96          & 68.20          & 59.71          & 62.51          & 65.17          \\
ProtT3 w/ MLP Proj.      & $\text{Galac}_{\text{1.3B}}$ & 4.07                 & 71.88           & 63.30            & 74.15            & 66.59          & 68.05          & 72.39          \\
ProtT3 w/o stage 1 & $\text{Galac}_{\text{1.3B}}$ & 4.85           & 72.33          & 64.79          & 75.67          & 68.21          & 69.34                & 73.00                \\
ProtT3   & $\text{Galac}_{\text{1.3B}}$ & \textbf{5.48} & \textbf{76.53} & \textbf{68.67} & \textbf{78.29} & \textbf{70.50} & \textbf{71.40} & \textbf{76.76} \\\bottomrule
\end{tabular}
\caption{Performances (\%) on the ProteinKG25~\cite{OntoProtein} dataset.}
\label{tab:proteinkg25}
\end{subtable}
\vspace{-2mm}
\caption{Protein captioning performances on Swiss-Prot and ProteinKG25. \textbf{Bold} indicates the best performance.}
\label{tab:caption}
\vspace{-4mm}
\end{table*}

\subsection{Stage 2: Protein-to-Text Generation Training}
\label{sec:generation}
In this stage, we train ProtT3 for protein-to-text generation. As Figure~\ref{fig:framework_arch} illustrates, we connect the cross-modal projector to the LM, feeding the protein representations $\Mat{Z}$ into the LM, so as to condition the text generation process by protein information. Note that, we use a linear layer to project $\Mat{Z}$ to the same dimension of the LM's input.

We train ProtT3 for each generation dataset separately, and append different text prompts after the protein representations to further control the generation process. For example, we use the text prompt of ``\textit{Describe this protein's function}'' for protein captioning, and ``\textit{Question: Does this protein contain polymer entities? Answer:}'' for protein QA. For training, we use the same loss as the protein captioning task in the previous section.

For training efficiency, we incorporate Low-Rank Adaptation (LoRA)~\cite{LoRA} adapters into the LM. LoRA adds pairs of trainable rank decomposition matrices into the selected weights of the LM. For example, LoRA modifies a pretrained weight $\Mat{W}_0\in \Space{R}^{d_1\times d_2}$ by adding a pair of matrices $\Mat{B}\in \Space{R}^{d_1\times r}$ and $\Mat{A}\in \Space{R}^{r\times d_2}$: 
\begin{equation}
    \Mat{W} = \Mat{W}_0 + \Mat{B}\Mat{A},
\end{equation}
where $\Mat{W}_0$ is kept frozen and only the newly added $\Mat{B}\Mat{A}$ is tuned. By using a small rank $r\ll \min(d_1, d_2)$, LoRA can adapt an LM to a new task while requiring little memory for storing gradients. This method has shown comparable performances to full-parameter fine-tuning~\cite{LoRA}.

\begin{table*}[t]
\small
\centering
\setlength{\tabcolsep}{4mm}
\begin{subtable}[t]{\textwidth}
\small
\centering
\begin{tabular}{lcccccccc}\toprule
                    & \multicolumn{4}{c}{Retrieval in batch}                        & \multicolumn{4}{c}{Retrieval in test set}                     \\ \cmidrule(lr){2-5} \cmidrule(lr){6-9}
                    & \multicolumn{2}{c}{P2T}  & \multicolumn{2}{c}{T2P}  & \multicolumn{2}{c}{P2T}  & \multicolumn{2}{c}{T2P}  \\ \cmidrule(lr){2-3} \cmidrule(lr){4-5} \cmidrule(lr){6-7} \cmidrule(lr){8-9}
Model             & Acc           & R@20          & Acc           & R@20          & Acc           & R@20          & Acc           & R@20          \\\midrule
% ProtST            & 79.1          & 99.6          & 71.2          & 99.2          & 9.7           & 47.5          & 14.8          & 59.8          \\
ProtST   & 81.4         & 99.4         & 79.6         & 99.4         & 15.9         & 56.6         & 17.8         & 63.5         \\
ProteinCLAP & 95.5         & 99.3         & 95.7         & 99.3         & 44.1         & 94.2         & 46.6         & 94.1        \\
% ProteinCLAP           & 95.5          & 99.3          & 95.5          & 99.3          & 43.6          & 93.5          & 45.4          & 93.4          \\
ProtT3 w/o PCap     & \underline{97.2}               & \textbf{99.9}              & \underline{97.1}          & \textbf{99.9}          & \underline{66.4}                & \underline{95.3}                & 66.2          & \underline{95.2}          \\
ProtT3 w/o PTM & 96.9          & 99.5          & 96.7          & 99.5          & 66.0          & 95.2          & \underline{66.5}          & 94.9          \\
ProtT3            & \textbf{97.7} & \textbf{99.9} & \textbf{97.5} & \textbf{99.9} & \textbf{68.3} & \textbf{96.0} & \textbf{68.1} & \textbf{95.8} \\\bottomrule
\end{tabular}
\vspace{-1mm}
\caption{Performances (\%)  on the Swiss-Prot~\cite{SwissProt} dataset.}
\end{subtable}
\begin{subtable}[t]{\textwidth}
\small
\centering
\begin{tabular}{lcccccccc} \toprule
                  & \multicolumn{4}{c}{Retrieval in batch}                        & \multicolumn{4}{c}{Retrieval in test set}                     \\ \cmidrule(lr){2-5} \cmidrule(lr){6-9}
                  & \multicolumn{2}{c}{P2T}  & \multicolumn{2}{c}{T2P}  & \multicolumn{2}{c}{P2T}  & \multicolumn{2}{c}{T2P}  \\\cmidrule(lr){2-3} \cmidrule(lr){4-5} \cmidrule(lr){6-7} \cmidrule(lr){8-9}
Model             & Acc           & R@20          & Acc           & R@20          & Acc           & R@20          & Acc           & R@20          \\ \midrule
% ProtST            & 66.6          & 98.3          & 67.8          & 97.2          & 2.5           & 37.6          & 3.6           & 40.4          \\
ProtST   & 70.8         & 98.5         & 70.9         & 98.2         & 5.5          & 41.6         & 5.8          & 43.3         \\
ProteinCLAP & 93.2         & 99.2         & 93.2         & 99.3         & 39.0         & 89.4         & 39.3         & 89.7        \\
% ProteinCLAP          & 92.7          & 99.3          & 93.2          & 99.4          & 34.7          & 88.2          & 36.7          & 88.5          \\
ProtT3 w/o PCap     & \textbf{95.1}               & \underline{99.8}              & \underline{95.0}          & \textbf{99.9}          & 53.4                & 91.2                & 53.0          & 91.2          \\
ProtT3 w/o PTM & 94.8          & 99.4          & 94.7          & 99.3          & \underline{53.8}          & \underline{91.3}          & \underline{54.1}          & \underline{91.3}          \\
ProtT3            & \textbf{95.1} & \textbf{99.9} & \textbf{95.3} & \textbf{99.9} & \textbf{55.8} & \textbf{91.7} & \textbf{55.6} & \textbf{91.7} \\ \bottomrule
\end{tabular}
\vspace{-1mm}
\caption{Performances (\%)  on the ProteinKG25~\cite{OntoProtein} dataset.}
\end{subtable}
\vspace{-3mm}
\caption{Protein-text retrieval performances. \textbf{Bold} indicates the best performance and \underline{underline} indicates the second best performance. We report performances of protein-to-text retrieval (P2T) and text-to-protein (T2P) retrieval. 
% Baseline performances are obtained by running their source codes~\cite{ProtST, ProteinDT}. 
}
\label{tab:retrieval}
\vspace{-4mm}
\end{table*}

\section{Experiments}
We begin by introducing the datasets of protein-text pairs, followed by the experimental results. Details on experimental settings, such as hyperparameters and baseline implementations, are provided in Appendix~\ref{app:exp}. In the experiments, ProtT3 utilizes $\text{Galactica}_{\text{1.3B}}$ as the base LM and $\text{ESM-2}_{\text{150M}}$ as the PLM.

% Throughout the experiments, ProtT3 employs $\text{Galactica}_{\text{1.3B}}$ as the base LM, and uses $\text{ESM-2}_{\text{150M}}$ as the PLM.

\subsection{Protein-Text Dataset Collection}
\label{sec:dataset}
In this section, we detail the protein-text pair datasets and the data processing procedure employed in our study. For all the used datasets, we discard protein sequences longer than $1022$ tokens, and carefully split the datasets to ensure no overlap between train/valid/test sets. Dataset statistics are available in Table~\ref{tab:statistics}, and details are in Appendix~\ref{app:data}.

% For captioning and retrieval, we use Swiss-Prot~\cite{SwissProt} and ProteinKG25~\cite{OntoProtein} datasets, whereas the PDB-QA~\cite{ProteinChat} dataset is used for QA. Dataset statistics are available in Table~\ref{tab:statistics}, and additional details can be found in Appendix~\ref{app:data}.

\textbf{Swiss-Prot}~\cite{SwissProt} is a protein sequence database with text annotations. We process the dataset following~\cite{ProtST}, but excluding protein names from the text annotations to prevent information leakage. The resulting text descriptions concatenate the annotations of protein functions, locations, and families.

\textbf{ProteinKG25}~\cite{OntoProtein} is a knowledge graph derived from the Gene Ontology~\cite{GeneOntology} database. We transform its triples into free texts by first aggregating the triples of the same protein, and then filling the protein information into a pre-defined text template. 

\textbf{PDB-QA}~\cite{ProteinChat} is a protein single-turn QA dataset derived from RCSB PDB\footnote{\url{https://www.rcsb.org}}. It includes 30 question templates about proteins' structures, properties, and supplementary information. As shown in Table~\ref{tab:sample}, to enable a fine-grained evaluation, we categorize the questions into four types, based on the answer's format (string or number) and content focus (structure/property or supplementary information).
It is worth noting that supplementary information is hard to predict given the protein sequence alone.
% ``structure/property'' refers to questions about proteins' structures or properties; and ``supplementary information'' refers to questions concerning proteins' supplementary information. 

\textbf{Training Pipeline.} In training stage 1, we train ProtT3 on the combination of the Swiss-Prot and ProteinKG25 datasets for protein-text retrieval. In training stage 2, we load the checkpoint from stage 1 and conduct separate fine-tuning on the three datasets for protein-to-text generation tasks.

\subsection{Protein Captioning}
We evaluate protein captioning performances on the Swiss-Prot and ProteinKG25 datasets. Following~\cite{MolT5}, we use the evaluation metrics of BLEU~\cite{BLEU}, ROUGE~\cite{ROUGE}, and METEOR~\cite{METEOR}. Additionally, we report the percentage that the prediction exactly matches the ground truth annotation. For comparison and ablation study purposes, we report the performances of LoRA fine-tuned $\text{Galactica}_{\text{1.3B}}$, and ProtT3's two variants: 1) ProtT3 w/ MLP Proj., which replaces ProtT3's cross-modal projector by an MLP, following~\cite{Llava-1.5}; and 2) ProtT3 w/o stage 1, which skips ProtT3's training stage 1. We do not compare with ProteinChat~\cite{ProteinChat} because it requires 3D protein structures, which are unavailable for these two datasets.

Table~\ref{tab:caption} presents the results. We observe the following: 1) ProtT3 and its variants substantially outperform the LoRA fine-tuned Galactica$_{\text{1.3B}}$, with ProtT3 showing a 10-point improvement in BLEU-2 scores. This underscores the significance of incorporating a PLM for protein understanding and ProtT3's effectiveness in undertanding protein inputs. 2) ProtT3 consistently surpasses its two variants on both datasets, highlighting the advantages of using a Q-Former projector and training stage 1.

\begin{table}[t]
\small
\centering
% \centering\resizebox{0.45\textwidth}{!}{
\setlength{\tabcolsep}{2pt}
\begin{tabular}{lcccccc} \toprule
                &                              & \multicolumn{2}{c}{String} & \multicolumn{2}{c}{Number} &         \\ \cmidrule(lr){3-4}  \cmidrule(lr){5-6} 
Model          & Base LM                      & SP      & SI      & SP      & SI      & Overall \\\midrule
% Random  guess  & -                            & 85.0         & 69.2        & 46.5         & 39.1        & 64.6    \\
LoRA ft         & $\text{Galac}_{\text{1.3B}}$ & 82.2                 & 65.7                 & 45.5                 & 38.0                 & 62.5                 \\
LoRA ft, Q-only & $\text{Galac}_{\text{1.3B}}$ & 76.0                 & 66.4                 & 44.7                 & \textbf{40.4}        & 60.2                 \\
ProteinChat     & $\text{Vicuna}_{\text{13B}}$ & 7.2                           & 10.3                         & 28.9                          & 19.2                         & 15.5         \\
ProtT3 w/o stage 1      & $\text{Galac}_{\text{1.3B}}$ & \underline{84.3}                          & \underline{67.8}                         & \underline{46.2}                          & 39.0                         & \underline{63.8}               \\
ProtT3          & $\text{Galac}_{\text{1.3B}}$ & \textbf{85.4}        & \textbf{69.8}        & \textbf{47.2}        & \underline{39.4}                 & \textbf{65.0}    \\\bottomrule
\end{tabular}
\vspace{-2mm}
\caption{Exact match performances (\%) for protein QA on the PDB-QA dataset~\cite{ProteinChat}. We categorize the QA pairs by their types. SP stands for structure/property, SI stands for supplementary information, and ft stands for fine-tuning. ProteinChat is evaluated using a checkpoint shared by the authors.}
\label{tab:qa}
\vspace{-5mm}
\end{table}

\begin{table*}[t]
\centering
\small
\begin{subtable}[t]{\textwidth}
\centering
\small
\begin{tabular}{lcccccccc} \toprule
\multicolumn{1}{l}{} & \multicolumn{4}{c}{Retrieval in batch}                        & \multicolumn{4}{c}{Retrieval in test set}                     \\ \cmidrule(lr){2-5} \cmidrule(lr){6-9}
\multicolumn{1}{l}{} & \multicolumn{2}{c}{P2T (\%)}   & \multicolumn{2}{c}{T2P (\%)}   & \multicolumn{2}{c}{P2T (\%)}   & \multicolumn{2}{c}{T2P (\%)}   \\ \cmidrule(lr){2-3} \cmidrule(lr){4-5} \cmidrule(lr){6-7} \cmidrule(lr){8-9}
PLM                  & Acc           & R@20          & Acc           & R@20          & Acc           & R@20          & Acc           & R@20          \\ \midrule
ESM-2$_{\text{8M}}$             & 96.6          & \textbf{99.9}          & 96.2          & 99.8          & 64.8          & 94.1          & 64.4          & 93.7          \\
ESM-2$_{\text{35M}}$            & 96.8          & \textbf{99.9}          & 96.7          & \textbf{99.9}          & 66.4          & 94.7          & 66.1          & 94.6          \\
ESM-2$_{\text{150M}}$           & \textbf{97.7} & \textbf{99.9} & \textbf{97.5} & \textbf{99.9} & \textbf{68.3} & \textbf{96.0} & \textbf{68.1} & \textbf{95.8} \\ \bottomrule
\end{tabular}
\caption{Performances (\%)  on the Swiss-Prot~\cite{SwissProt} dataset.}
\end{subtable}
\begin{subtable}[t]{\textwidth}
\centering
\small
\begin{tabular}{lcccccccc} \toprule
\multicolumn{1}{l}{} & \multicolumn{4}{c}{Retrieval in batch}                        & \multicolumn{4}{c}{Retrieval in test set}                     \\
\multicolumn{1}{l}{} & \multicolumn{2}{c}{P2T (\%)}   & \multicolumn{2}{c}{T2P (\%)}   & \multicolumn{2}{c}{P2T (\%)}   & \multicolumn{2}{c}{T2P (\%)}   \\ 
PLM                  & Acc           & R@20          & Acc           & R@20          & Acc           & R@20          & Acc           & R@20          \\ \midrule
ESM-2$_{\text{8M}}$             & 93.1          & 99.8          & 93.5          & 99.7          & 51.9          & 88.6          & 51.7          & 89.0          \\
ESM-2$_{\text{35M}}$            & 94.0          & 99.8          & 94.3          & 99.8          & 54.1          & 89.8          & 54.0          & 90.1          \\
ESM-2$_{\text{150M}}$           & \textbf{95.1} & \textbf{99.9} & \textbf{95.3} & \textbf{99.9} & \textbf{55.8} & \textbf{91.7} & \textbf{55.6} & \textbf{91.7} \\ \bottomrule
\end{tabular}
\caption{Performances (\%)  on the ProteinKG25~\cite{OntoProtein} dataset.}
\end{subtable}
\caption{Ablation studies of PLMs for molecule-text retrieval.}
\label{tab:ablate_stage1}
\end{table*}

\begin{table*}[t]
\small
\centering
\setlength{\tabcolsep}{2pt}
\begin{tabular}{lc|ccccccc} \toprule
Base LM                        & Exact Match   & BLEU-2         & BLEU-4         & ROUGE-1        & ROUGE-2        & ROUGE-L        & METEOR         \\ \midrule
$\text{Phi-1.5}_{\text{1.3B}}$ & 1.13          & 71.71          & 63.65          & 68.62          & 60.29          & 61.52          & 69.28          \\ 
$\text{Galac}_{\text{1.3B}}$   & \textbf{5.48} & \textbf{76.53} & \textbf{68.67} & \textbf{78.29} & \textbf{70.50} & \textbf{71.40} & \textbf{76.76} \\ \bottomrule
\end{tabular}
\caption{Ablation studies of textual LMs for protein captioning on the ProteinKG25~\cite{OntoProtein} dataset.} \label{tab:ablate_stage2}
\end{table*}

\begin{figure*}
\centering
\small
\includegraphics[width=0.98\textwidth]{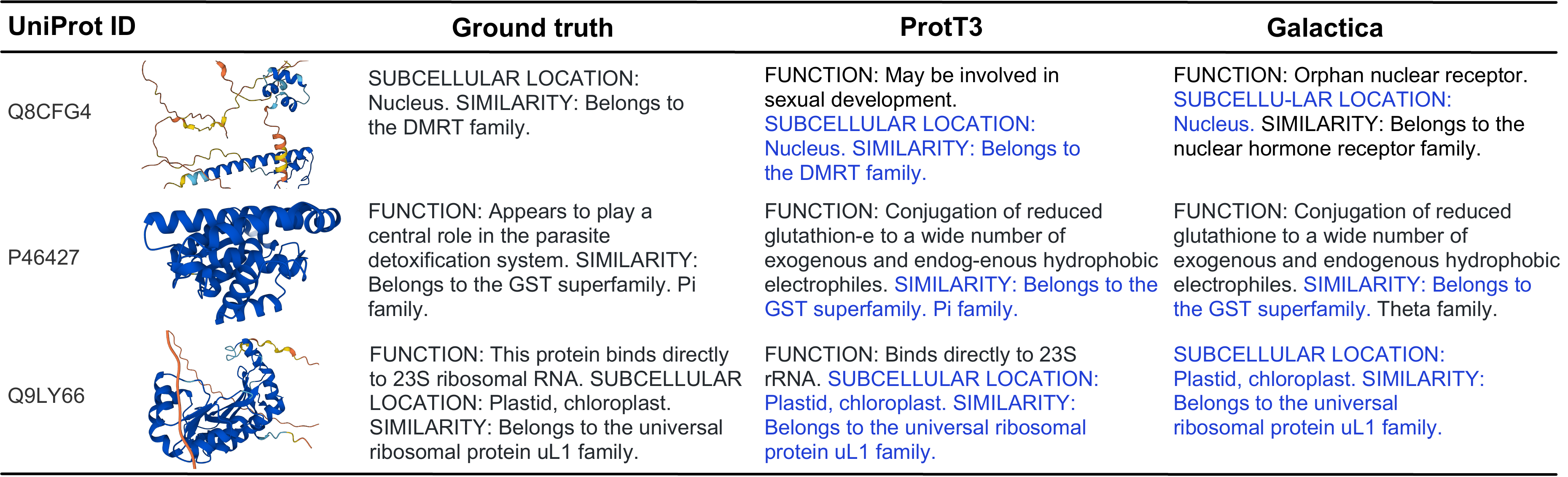}
\vspace{-1mm}
\caption{Protein captioning examples from Swiss-Prot. We highlight sentences that exactly match the ground truth. Figures of protein structures are generated by AlphaFold2~\cite{AlphaFold2}.}
\vspace{-2mm}
\label{fig:case_study}
% \vspace{-2mm}
\end{figure*}

\begin{figure}
\centering
\small
\includegraphics[width=0.47\textwidth]{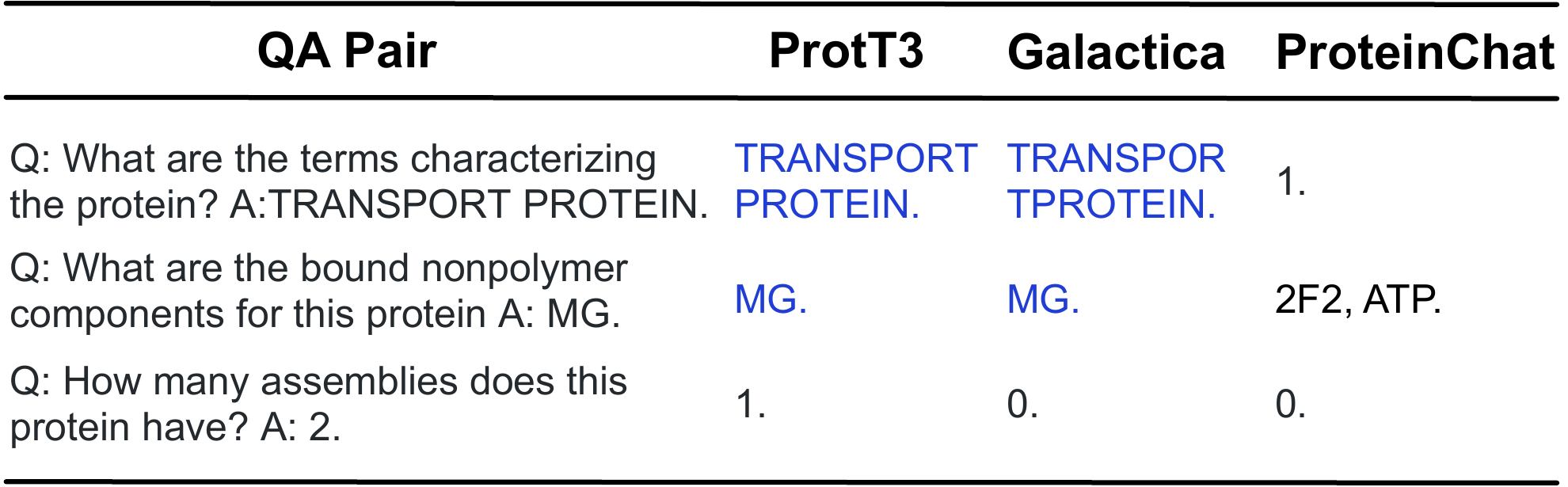}
% \vspace{-2mm}
\caption{Examples of protein QA results in the PDB-QA dataset. We highlight the correct predictions.}
\label{fig:qa_case_study}
\vspace{-3mm}
\end{figure}

\vspace{-3mm}
\subsection{Protein-Text Retrieval}
\vspace{-1mm}
We evaluate protein-text retrieval performances on the Swiss-Prot and ProteinKG25 datasets. Following~\cite{MoMu}, our evaluation includes protein-text retrieval in a batch of $64$ random samples and retrieval in the entire test set. We use Accuracy and Recall@20 as the evaluation metrics. For retrieval, ProtT3 first retrieves the top $128$ candidates by PTC, and then uses PTM for re-ranking. For comparison, we employ ProtST~\cite{ProtST} and ProteinCLAP~\cite{ProteinDT} as baselines, and present ProtT3 w/o PTM and ProtT3 w/o PCap for ablation studies. 

Table~\ref{tab:retrieval} shows the results. We observe that: 1) ProtT3 outperforms baselines by over $14\%$ accuracy for retrieval in the test set, highlighting its capability in aligning proteins with corresponding text descriptions. 2) PTM improves ProtT3's retrieval accuracy in test set by $1\sim 2\%$ on both datasets. This is because PTM allows protein and text information to interact at the Q-Former's early layers, achieving more fine-grained protein-text similarity measurements. 3) PCap improves ProtT3's retrieval accuracy by $\sim 2\%$. This is because 
PCap encourages the query tokens to extract protein information most 
relevant to the text input, therefore aiding in protein-text alignment.

\subsection{Protein Question-Answering}
We evaluate protein QA performances on the PDB-QA dataset. Considering that the answers in PDB-QA typically consist of $1\sim 2$ words, we select exact match as the evaluation metric. For comparison, we employ ProteinChat~\cite{ProteinChat} and LoRA fine-tuned $\text{Galactica}_{\text{1.3B}}$ as baselines. We also assess a $\text{Galactica}_{\text{1.3B}}$ that consumes only questions without proteins during training and prediction (\ie LoRA ft, Q-only). This baseline measures the dataset's uni-modal bias~\cite{RUBI}: the proportion of questions that can be correctly answered without looking at proteins. 

The results are shown in Table~\ref{tab:qa}. We observe that: 1) ProtT3 surpasses baselines by 2.5\% overall, and consistently outperforms them in predicting protein structures and properties. This demonstrates ProtT3's multi-modal understanding ability for the proteins and the textual questions. 
2) The baseline consuming question only (\ie LoRA ft, Q-only) presents comparable performance, indicating that the dataset has a substantial uni-modal bias. This finding suggests an opportunity to explore debias techniques~\cite{DBLP:conf/acl/MahabadiBH20} in future studies.
3) ProteinChat underperforms other methods, possibly because it only trains a linear layer between the frozen PLM and LM. The linear layer is insufficient to map the protein representations into the text space of the LM, and adapt the LM to QA.
4) Models perform worse on questions about numbers or supplementary information. This observation suggests the potential benefit of augmenting these models with external tools~\cite{toolformer} like calculators and search engines.

\subsection{Ablation Studies on Pretrained Models}
Here we conduct ablation studies on the pretrained models used in our method. Specifically, we ablate the impact of different PLMs in training stage 1, and ablate different LMs in training stage 2.

\textbf{Ablating PLMs.} In training stage 1, we replace the ESM-2$_{\text{150M}}$ protein encoder with its smaller variants, namely ESM-2$_{\text{8M}}$ and ESM-2$_{\text{35M}}$, to evaluate their performances for protein-text retrieval. As shown in Table~\ref{tab:ablate_stage1}, we can observe that retrieval performance increases monotonically with model size, a trend consistent with previous observations in the LM domain.

\textbf{Ablating LMs.} In training stage 2, we replace Galactica$_{\text{1.3B}}$ to Phi-1.5$_{\text{1.3B}}$~\cite{li2023textbooks}. Unlike Galactica, Phi-1.5$_{\text{1.3B}}$ is pretrained on general domain data but not focusing on scientific literature. Table~\ref{tab:ablate_stage2} shows protein captioning performance on the ProteinKG25~\cite{OntoProtein} dataset. We can observe that Galactica$_{\text{1.3B}}$ significantly outperforms Phi-1.5$_{\text{1.3B}}$ for protein captioning, although they have similar sizes. We attribute this performance gap to their pretraining corpus, with Galactica$_{\text{1.3B}}$ performing better for pretraining on more scientific literature.

\subsection{Examples of Protein-to-Text Generation}
% Figure~\ref{fig:case_study} shows three examples to compare the protein captioning results between ProtT3 and a LoRA fine-tuned Galactica. 
Figure~\ref{fig:case_study} shows three examples of protein captioning. 
In the first example, ProtT3's caption is more accurate by correctly identifying the DMRT family, while Galactica does not. In the second example, both models fail to identify the protein's function. Nevertheless, ProtT3's prediction regarding the protein family is more accurate. The third example shows that both models successfully predict the subcellular location and protein family. ProtT3 goes a step further by predicting the protein's function, which is closer to the ground truth description.

Figure~\ref{fig:qa_case_study} shows three examples of protein QA. We can observe that both ProtT3 and Galactica answer the first two questions about protein property/structure correctly, and fail on the third question, which requires a numerical answer. On the other hand, ProteinChat struggles with all three questions, failing to answer each of them.

\section{Conclusion and Future Works}
In this work, we propose ProtT3, a new protein-text modeling framework. ProtT3 aims to facilitate text-based protein understanding via protein-to-text generation and protein-text retrieval. To achieve this, ProtT3 integrates a PLM into an LM to enhance the LM's protein understanding ability. A cross-modal projector enables this integration by bridging the modality gap between the two modules. To promote future research, we set benchmarks for protein-text modeling tasks, including protein captioning, protein QA, and protein-text retrieval, where ProtT3 significantly outperforms existing baselines. 

Looking ahead, we plan to explore enabling LMs to understand 3D protein structures~\cite{3D-MoLM} and apply this understanding to more tasks of drug discovery, property prediction~\cite{liu2023rethinking,li2022let,li2023selfattentive}, molecule generation~\cite{luo2024textguided}, and OOD generalization~\cite{fangjf2023eva, fangjf2023exgc}.

\section*{Limitations}
% \subsection{}
\textbf{On Representing Proteins by 1D Sequences.} Proteins can be represented by either 1D sequences of amino acids or 3D coordinates of their atoms. In this study, we represent proteins by 1D sequences considering their abundance compared to 3D structures. As of 2023, the Protein Data Bank contains 220K 3D structures~\footnote{\url{https://www.wwpdb.org/stats/deposition}}, whereas the UniProt database houses 227M 1D sequences~\cite{uniprot2023uniprot}. 
This abundance of 1D sequences enables us to collect a larger dataset. In Section~\ref{sec:dataset}, we collect a total of 1M protein-text pairs, a significant increase from the 143K pairs in the earlier work~\cite{ProteinChat} using 3D structures.

While it is feasible to predict the 3D structures of protein sequences by running protein folding algorithms~\cite{AlphaFold2,ESM2}, this prediction process is notably time-consuming. For example, predicting the 3D structures of the 440k protein sequences in the ProteinKG25~\cite{OntoProtein} dataset would take approximately four months when using an A100 80GB GPU. Due to this computational demand, we choose to focus on 1D sequences, leaving the exploration of 3D structures for future research.

\textbf{On Dynamic Protein Structures.} Proteins have dynamic structures, and natural language is limited in fully depicting protein dynamics.
% , and protein-to-text generation cannot fully reveal the properties and structures of proteins.
Nonetheless, similar to video-to-text generation~\cite{DBLP:journals/csur/AafaqMLGS20}, where language partially describes dynamic videos, our work on protein-to-text generation is a crucial step towards encapsulating complex biological phenomena in a human-readable form.

\section*{Potential Ethics Impact}
In this study, the proposed method and dataset focus on proteins and their properties, and include no human subjects. Consequently, we believe this study presents no direct ethical concerns. 
However, the proposed LMs can be abused to generate biased or toxic information, and can generate inaccurate description of protein properties and structures. Therefore, the ethical implications of our work align with those common to LM research.

\section*{Acknowledgement}
This research is supported by the National Science and Technology Major Project (2023ZD0121102), National Natural Science Foundation of China (92270114). This research is partially supported by the National Research Foundation Singapore under the AI Singapore Programme (AISG Award No: AISG2-TC-2023-010-SGIL), the Singapore Ministry of Education Academic Research Fund Tier 1 (Award No: T1 251RES2207) and the Google Cloud Research Credits program with the award (Q4MJ-YH1K-3MVX-FP6Q). This research is supported by NExT Research Center.

\bibliography{reference}

\appendix

\clearpage
\newpage
\pagebreak

\begin{table*}[t]
\small
\centering
\begin{tabular}{p{8.5cm}|l|p{2.5cm}}\toprule
\textbf{Question}                                                                                                        & \textbf{Accuracy} & \textbf{Number of candidate answers} \\ \midrule
Does this protein contain RNA polymer   entities?                                                               & 0.9871   & 2                           \\
Does this protein contain DNA polymer   entities?                                                               & 0.9790   & 2                           \\
How many nucleic acid polymer entities(DNA or   RNA) does this protein have?                                    & 0.9644   & 7                           \\
Does this protein contain branched entities?                                                                    & 0.9509   & 2                           \\
What is the polymer entity type for this   protein?                                                             & 0.9192   & 4                           \\
How many model structures deposited for this   protein?                                                         & 0.9036   & 39                          \\
What are the software programs reported in   connection with the production of this protein?                    & 0.0733   & 3611                        \\
What is the molecular mass (KDa) of polymer   and non-polymer entities (exclusive of solvent) for this protein? & 0.0037   & 7367                        \\
How many heavy atom coordinates records does   this protein have?                                               & 0.0010   & 6550         \\ \bottomrule              
\end{tabular}
\caption{Sampled questions of structures/properties and the corresponding accuracies and numbers of candidate answers.} \label{tab:error1}
\end{table*}

\begin{table*}[t]
\centering
\small
\begin{tabular}{p{4.5cm}|c} \toprule
\textbf{Question} & \textbf{Candidate answers}                                       \\ \midrule
How many model structures deposited for this protein?                             & \{1, 2, 3, 4, 5, 6, 7, 8, 9, 10, 11, 12, 13, ...\}                 \\
What experimental method(s) were used to determine the structure of this protein? & \{'NMR', 'X-ray', 'EM', 'Other', 'Multiple methods', 'Neutron'\}   \\
When is this protein first published?                                             & \{1975, 1976, 1977, 1979, 1980, 1981, 1982, 1983, 1984, 1985 ...\} \\ \bottomrule
\end{tabular}
\caption{Sampled questions of supplementary information and the candidate answers.}\label{tab:error2}
\end{table*}

\begin{table*}[t!]
\small
\centering
\begin{tabular}{lp{12cm}} \toprule
\textbf{Entry Name} & \textbf{Description}                                                                                                                                                                                                                                                                                                                                                                                                                                                                                                   \\ \midrule
ENO1\_CHLTE & FUNCTION: Catalyzes the reversible conversion of 2-phosphoglycerate into phosphoenolpyruvate. It is essential for the degradation of carbohydrates via glycolysis. SUBCELLULAR LOCATION: Cytoplasm Secreted Cell surface   Note=Fractions of enolase are present in both the cytoplasm and on the cell surface. The export of enolase possibly depends on the covalent binding to the substrate; once secreted, it remains attached to the cell surface.   SIMILARITY: Belongs to the enolase family. \\ \midrule
6PGD\_AGGAC & FUNCTION: Catalyzes the oxidative decarboxylation of 6-phosphogluconate   to ribulose 5-phosphate and CO(2), with concomitant reduction of NADP to   NADPH. SIMILARITY: Belongs to the 6-phosphogluconate dehydrogenase family.                                                                                                                                                                                                                                                                               \\ \midrule
RIMP\_ACAM1 & FUNCTION: Required for maturation of 30S ribosomal subunits. SUBCELLULAR   LOCATION: Cytoplasm. SIMILARITY: Belongs to the RimP family. \\ \bottomrule
\end{tabular}
\caption{Examples of protein descriptions in the Swiss-Prot dataset. Entry name is the protein index in the original database.}
\label{tab:swissprot_examples}
\end{table*}
\section{Details of Protein-Text Dataset Collection}
\label{app:data}
\textbf{Swiss-Prot}~\cite{SwissProt}. Dataset samples are provided in Table~\ref{tab:swissprot_examples}. We largely follow the instructions in~\cite{ProtST} for data processing. We first download the lastest dump of Swiss-Prot from the UniProt website\footnote{\url{https://www.uniprot.org/}}, and then select three annotation fields: \textit{Function}, which describes protein's roles such as catalysis and transport; \textit{Subcellular Location}, indicating where a protein is typically found within a cell; \textit{Similarity}, which details the protein's family. These three annotation fields are then concatenated to form the text description of proteins. Different from~\cite{ProtST}, we do not include the annotation of \textit{Protein Name} in the text description. This is because \textit{Protein Name} can be potential information leakage for the protein-text retrieval task. 

\textbf{ProteinKG25}~\cite{OntoProtein}. Dataset samples are shown in Table~\ref{tab:proteinkg25_examples}. ProteinKG25 is originally a knowledge graph derived from the Gene Ontology database~\cite{GeneOntology}. ProteinKG25 is stored as triples in the format of (protein sequence, relation, property), such as \textit{(protein 1, enables, structural constituent of ribosome)}, \textit{(protein 2, is involved in metabolic process, rRNA processing)}, and \textit{(protein 3, is located in, cytoplasm)}. To synthetic text descriptions using such triples, we perform the following steps: 
\begin{itemize}[leftmargin=*]
\item Group triples by protein sequences. Each resulting group has triples of the same protein.
\item Further group triples by relation types. Each resulting group has triples of the same protein and the same relation type. For example, a group can include triples of \textit{(protein 1, is involved in lipid metabolic process, fatty acid biosynthetic process)}, \textit{(protein 1, is involved in lipid metabolic process, lipid catabolic process)}, and \textit{(protein 1, is involved in lipid metabolic process, fatty acid metabolic process)}.
\item Transform the triples in each group into a sentence by filling the information into a text template. For example, the three triples in the previous step will be transformed into: \textit{This protein is involved in the following lipid metabolic process: fatty acid biosynthetic process, lipid catabolic process, and fatty acid metabolic process.} 
\item Concatenate the sentences of the same protein together to form the final text description.
\end{itemize}
Note that, we do not include the protein sequence information in the text description to avoid information leakage. The Python script for data processing and the text templates will be made available online to facilitate future research.

\textbf{PDB-QA}~\cite{ProteinChat}. The categorization of all 30 question templates is detailed in Table~\ref{tab:pdb_cate}. The original PDB-QA dataset provides only 3D structures of proteins. To retrieve their 1D sequences, we call the web API of the RCSB PDB website\footnote{\url{https://www.rcsb.org/}} using the proteins' PDB IDs.

\begin{table*}[t]
\centering
\small
\begin{tabular}{lclc}\toprule
\multicolumn{2}{c}{Retrieval   Task}             & \multicolumn{2}{c}{Generation Task}                         \\ \cmidrule(lr){1-2} \cmidrule(lr){3-4}
Model          & \multicolumn{1}{l}{\#Parameters} & Model                     & \multicolumn{1}{l}{\#Parameters} \\ \midrule 
ProtST         & 786M                            & ProteinChat               & 13.6B                           \\
ProteinCLAP    & 530M                            & Galactica$_{\text{1.3B}}$ & 1.3B                            \\
ProtT3 stage 1 & 326M                            & ProtT3 stage 2            & 1.6B                           \\ \bottomrule
\end{tabular}
\caption{Number of parameters of ProtT3 and key baselines.}
\label{tab:para_number}
\end{table*}

\section{Error Analysis}
\textbf{Questions of Structures/Properties.} Table~\ref{tab:error1} presents the statistics of sampled questions in this category. We can observe a strong correlation between accuracy and the number of candidate answers. This observation is intuitive, as questions become more challenging with an expanded answer pool. Further, questions that have a large number of candidate answers are those that require numerical outputs. This highlights the potential of improving the ability of counting substructures and processing numbers in future studies.

\textbf{Questions of Supplementary Information.} Table~\ref{tab:error2} shows the sampled questions of supplementary information and the candidate answers. We can observe that the questions and answers are not directly relevant to protein structures or functions. Therefore, we infer that the success of these questions often relies on exploiting the dataset's inherent biases rather than understanding protein structures and properties. This realization prompts a concern that the model's performance could significantly diminish when applied to datasets without similar biases, indicating a potential area for improvement using retrieval augmentation.

\section{Experimental Details}
\label{app:exp}
Following MolT5~\cite{MolT5} and ProteinCLAP~\cite{ProteinDT}, we use a single same random seed for experiments. Using multiple random seeds is cost-prohibitive for tuning large LMs.

The number of parameters of ProtT3 and keybaseline are summarized in Table~\ref{tab:para_number}. All the experiments are conducted on either two A100-40GB GPUs or four V100-32GB GPUs. We have used FlashAttention-2~\cite{FlashAttention2} to speedup the ESM-2 and Galactica model.

\subsection{Experimental Details of ProtT3}
\textbf{Hyperparameters.} The training stage 1 has 50 epochs and stage 2 has 10 epochs. The batch size is 128 for both stage 1 and stage 2.
Q-Former has 8 query tokens ($N_q=8$). The optimizer is configured following~\cite{BLIP2}. We use an AdamW optimizer and a learning rate scheduler that is a combination of linear warmup and cosine decay. The peak learning rate is $1e-4$ and the warmup has $1000$ steps. Weight-decay is set to $0.05$. For all experiments that involve fine-tuning the $\text{Galactica}_{\text{1.3B}}$ LM, we set LoRA's rank $r$ to $8$ and apply LoRA to Galactica's weights of \texttt{[q\_proj, v\_proj, k\_proj, out\_proj, fc1, fc2]}. This configuration yields LoRA adapters with 7M parameters, constituting merely $0.54\%$ of the parameters in the $\text{Galactica}_{\text{1.3B}}$. LoRA is implemented using the OpenDelta~\cite{OpenDelta} library. For the text generation training, we truncate all the texts to at most 128 tokens.

\textbf{Case Studies.} The examples in Figure~\ref{fig:case_study} are sampled from the predictions of models in Table~\ref{tab:swissprot}. The examples in Figure~\ref{fig:qa_case_study} are sampled from the predictions of models in Table~\ref{tab:proteinkg25}. Specifically, the Galactica model in Figure~\ref{fig:case_study} and Figure~\ref{fig:qa_case_study} refers to the LoRA fine-tuned Galactica on the corresponding dataset.

\subsection{Experimental Details of Baselines and ProtT3 Variants.}

\textbf{LoRA fine-tune Galactica$_{\text{1.3B}}$}. We use LoRA fine-tune Galactica$_{\text{1.3B}}$ as a baseline to verify the effectiveness of using a PLM for protein understanding. The hyperparameters are the same as ProtT3: they use the same optimizer, learning rate, batch size, epochs, and LoRA configuration. This baseline is applied directly for training stage 2  without training stage 1 because of its incapability of protein-text retrieval.

\textbf{ProtT3 w/ MLP Proj}. This ProtT3 variant replaces ProtT3's Q-Former projector with a two-layer MLP following~\cite{Llava-1.5}. Hyperparameters are the same as ProtT3. This variant includes training stage 2 without stage 1, because the MLP projector cannot process multi-modal inputs, therefore cannot be applied in stage 1. 

\textbf{ProtT3 w/o Stage 1}. This ProtT3 variant is to verify the effectiveness of training stage 1 for downstream generation tasks. Hyperparameters are the same as ProtT3.

\textbf{ProtT3 w/o PTM}. This is a variant of ProtT3's stage 1 model that removes the PTM objective function. Without the PTM objective, ProtT3 relies on retrieval ability of the PTC objective for protein-text retrieval.

\textbf{ProtT3 w/o PCap}. This is a variant of ProtT3's stage 1 model that removes the PCap objective function.

\textbf{ProtST}~\cite{ProtST}. We re-implement their method on our datasets, using
their released codes and the original hyperparameters in their paper. Specifically, we have implemented their ProtST-ESM-2 model and their multi-modal pretraining stage. Following their original paper, we adopt ESM-2$_{\text{650M}}$ as the protein encoder and adopt PubMedBERT-abs$_{\text{109M}}$~\cite{PubMedBERT} as the text encoder. 
The model is trained on the combination of the Swiss-Prot and ProteinKG25 datasets, which is the same as ProtT3's training stage 1. The pretrained model is then applied for protein-text retrieval directly.

\textbf{ProteinCLAP}~\cite{ProteinDT}. We re-implement their method from scratch for their source codes are not publicly available. Our implementation largely follows the description in their paper: using SciBERT~\cite{SciBERT} for text encoding and ProtBERT~\cite{ProtTrans} for protein encoding, learning rate $1e-5$. We have trained the model on the combination of the Swiss-Prot and ProteinKG25 datasets for 50 epochs, which is the same as ProtT3's training stage 1. The pretrained model is then applied for protein-text retrieval.

\textbf{ProteinChat.} ProteinChat~\cite{ProteinChat} is developed using the base LM of $\text{Vicuna}_{\text{13B}}$~\cite{Vicuna}. We did not evaluate ProteinChat for the Swiss-Prot and ProteinKG25 datasets because these two datasets do not include proteins' 3D structures, which is required by ProteinChat's protein encoder. For the PDB-QA dataset, we evaluate a checkpoint shared by the authors, which is the checkpoint used in their original paper.

\begin{table*}[t!]
\centering
\small
\begin{tabular}{lp{12cm}}\toprule
\textbf{Protein ID} & \textbf{Description}                                                                                                                                                                                                                                                                                                                                                                                                                                                                               \\\midrule
B1VNN9     & This protein is involved in the following process: nitrogen compound metabolic process. This protein is located in the following component:   cytoplasm. This protein is involved in metabolic process: nitrogen compound metabolic process. This protein enables metal ion binding: nickel cation binding. This protein enables the following function: nickel cation binding.                                                                                                     \\\midrule
Q01644     & This protein is part of the following component: cellular component. This protein is involved in the following process: multicellular organism development, cell differentiation, sperm axoneme assembly, and   spermatogenesis. This protein enables the following function: molecular function.                                                                                                                                                                                  \\\midrule
Q32AB9     & This protein is involved in metabolic process: tricarboxylic acid cycle,   carbon fixation, and oxaloacetate metabolic process. This protein enables metal ion binding: magnesium ion binding. This protein enables catalytic activity: hosphoenolpyruvate carboxylase activity, catalytic activity, and lyase activity. This protein enables the following functions: lyase activity,   magnesium ion binding, catalytic activity, and phosphoenolpyruvate carboxylase activity. \\\bottomrule
\end{tabular}
\caption{Examples of protein descriptions in the ProteinKG25 dataset. Protein ID is the protein index in the original dataset.}
\label{tab:proteinkg25_examples}
\end{table*}

\begin{table*}[t!]
\small
\centering
\begin{tabular}{p{4cm}p{2.5cm}|p{4.5cm}p{2.5cm}}\toprule
\textbf{Question}                                                             & \textbf{Type}             & \textbf{Question}                                                                                               & \textbf{Type}             \\\midrule
What is the nucleic acid polymer entity   type for this protein?              & String structure/property & How many heavy atom coordinates records does this protein have?                                                 & Number supplementary information   \\\midrule
When is this protein first published?                                         & Number supplementary information   & How many intermolecular metalic bonds does this protein have?                                                   & Number structure/property \\\midrule
How many polymer monomers does this   protein have?                           & Number structure/property & Does this protein have hybrid nucleic acid polymer entities?                                                    & String structure/property \\\midrule
How many assemblies does this protein   have?                                 & Number structure/property & What is the molecular mass (KDa) of polymer and non-polymer entities   (exclusive of solvent) for this protein? & Number structure/property \\\midrule
How many heavy solvent atom coordinates   records does this protein have?     & Number supplementary information   & Is this protein determined by experimental or computational methods?                                            & String supplementary information   \\\midrule
Does this protein have cis-peptide   linkages?                                & String structure/property & What is the radiation wavelength in angstroms for this protein?                                                 & Number structure/property \\\midrule
Does this protein contain branched   entities?                                & String structure/property & Does this protein have unmodeled polymer monomers?                                                              & String supplementary information   \\\midrule
How many entities does this protein have?                                     & Number structure/property & How many intermolecular covalent bonds does this protein have?                                                  & Number structure/property \\\midrule
Does this protein contain solvent   entities?                                 & String structure/property & How many model structures deposited for this protein?                                                           & Number supplementary information   \\\midrule
What is the polymer entity type for this   protein?                           & String structure/property & What are the software programs reported in connection with the production   of this protein?                    & String supplementary information   \\\midrule
How many nucleic acid polymer entities   (DNA or RNA) does this protein have? & Number structure/property & How many hydrogen atom coordinates records does this protein have?                                              & Number supplementary information   \\\midrule
What is the polymer entity composition   for this protein?                    & String structure/property & What experimental method(s) were used to determine the structure of this   protein?                             & String supplementary information   \\\midrule
Does this protein contain DNA polymer   entities?                             & String structure/property & Does this protein contain polymer entities?                                                                     & String structure/property \\\midrule
Does this protein contain non-polymer   entities?                             & String structure/property & Does this protein contain RNA polymer entities?                                                                 & String structure/property \\\midrule
What are the terms characterizing the   protein?                              & String structure/property & What are the bound nonpolymer components for this protein                                                       & String structure/property \\\bottomrule
\end{tabular}
\caption{Categorization of the 30 questions in the PDB-QA dataset~\cite{ProteinChat}.}
\label{tab:pdb_cate}
\end{table*}

\section{Further Discussions}
\subsection{Intuitions on Q-Former's Capabilities}
For clarity, we briefly summarize Q-Former's capability in the two stages:

\noindent \textbf{Stage 1: Protein-Text Retrieval Training.}
\begin{itemize}[leftmargin=*]
\item \textbf{Protein-Text Retrieval with Protein-Text Contrasting (PTC) and Protein-Text Matching (PTM).} Both of these two pretraining tasks are designed for retrieval: PTC is faster for using cosine similarity to model sequence-level similarity between proteins and texts; PTM is slower but more accurate for using self-attention to model the token-level similarity. The other pretraining task, Protein Captioning, although cannot be applied for retrieval directly, improves Q-Former's representation learning ability, so as to improve the retrieval performance.
\item \textbf{Warmup before stage 2.} The three pretraining tasks pretrain Q-Former to have multi-modal representation ability of texts and proteins, acting as a warmup training before stage 2.
\end{itemize}

\noindent \textbf{Stage 2: Protein-to-Text Generation Training.}
\begin{itemize}[leftmargin=*]
\item \textbf{Nonlinear Mapping from Protein Space to Text Space.} Q-Former maps the protein encoder's output to the input space of an LM. This innovative approach allows the LM to leverage pre-existing knowledge from the protein encoder, facilitating a deeper understanding of protein structures and functions.
\end{itemize}

\subsection{Discussion on Linear Mapping}
Proteins and texts inherently exist within high-dimensional spaces, exhibiting intricate and nonlinear semantic relationships. The utilization of a mere linear layer to map between protein and text domains often proves insufficient, potentially resulting in a loss of critical structural details of proteins or inaccuracies in text representatio. This claim is supported by the observation in Table~\ref{tab:qa}, where a baseline using a linear projector significantly underpforms the others.

It is important to note, however, that applying a linear layer between the protein encoder and the LM does not necessarily produce a strictly linear mapping from proteins to texts. Should either the protein encoder or the LM undergo fine-tuning — as opposed to remaining frozen — their weights can adapt to facilitate a nonlinear mapping between protein and text. Thus, our critique is not directed at the use of a linear layer per se, but rather at its employment in conjunction with frozen protein encoders and LMs.

\section{Discussion on Licensing}
\textbf{Pretrained Models and Codes.} The Galactica~\cite{Galactica} model is under the CC BY-NC 4.0 license, ESM-2~\cite{ESM2} under the MIT License, and Sci-BERT~\cite{SciBERT} under the Apache License. Collectively, these licenses permit non-commercial use and redistribution of the models.

\textbf{Datasets.} The Swiss-Prot~\cite{SwissProt} dataset is distributed under the CC BY 4.0 License. ProteinKG25~\cite{OntoProtein} dataset is released under the MIT License, its original source GeneOntology~\cite{GeneOntology} is under the CC BY 4.0 License. The PDB-QA~\cite{ProteinChat} dataset is released under the BSD 3-Clause License, its original source RCSB-PDB is under the CC0 1.0 Universal License. These licenses collectively allow the use and redistribution of the datasets.

\textbf{Our Licensing Approach.} In line with the licensing terms of the codes, models, and datasets we utilized, we will release our pretrained models and codes under the CC BY-NC 4.0 License. For our datasets, we will apply the CC BY 4.0 License.

\end{document}